\documentclass{article}
\usepackage{graphicx}
\usepackage{epsfig}
\usepackage{subeqn}
\begin{document}
\title{\bf Bose-Einstein correlations at LEP and Tevatron energies.}
\author {G.A.Kozlov$^2$, L.Lovas$^1$, S.Tokar$^1$, Yu.A.Boudagov$^2$, A.N.Sissakian$^2$}
\maketitle
\begin{itemize}
\item[1] {Comenius University, Faculty of mathematics, physics and informatics, Department of nuclear physics, Mlynska Dolina F1, 842 48 Bratislava, Slovakia}
\item[2] {Joint Institute for Nuclear Research, Joliot-Curie st. 6, 141980 Dubna, Russian Federation}
\end{itemize}

\section*{Abstract}
\indent Using the Bose-Einstein correlations (BEC) implemented in PYTHIA we investigated a possibility of the CDF experiment at the Tevatron to see the two-particle correlations in the final state of interactions. The approach based on quantum field theory at finite temperature was applied to the ALEPH data at LEP, and the BEC important parameters were retrieved.\\
\section{Introduction}
\indent Over the past few decades, a considerable number of studies have been done on the phenomena of multi-particle correlations induced by collisions between elementary particles. It is understood that the studies of correlations between produced particles, the effects of coherence and chaoticity, an estimation of particle emitting source size play an important role in high energy physics.\\
In papers [1,2], there were established and developed the master equations in the form of the field operator evolution equation (Langevin-like [3]), which allows one to gain a better understanding of the structure of the emitting source.\\
The shapes of Bose-Einstein correlation (BEC) function were established in the LEP experiments ALEPH $[4]$, DELPHI $[5]$ and OPAL $[6]$, and ZEUS Collaboration at HERA $[7]$, which also indicated a dependence of the measured correlation radius on the hadron $(\pi,K)$ mass.\\
There is still no definite explanation of the above mentioned mass dependence. The aim of this paper is, firstly, to carry out the numerical analysis applied to the LEP data (e.g., ALEPH Collaboration) on two-particle BEC in the framework of quantum field theory at finite temperature $(QFT_\beta)$ approach [1,2], and secondly, to give the proposal on BEC studies at the Tevatron energies.\\
\noindent We simulated $p \bar p$ annihilations at the Tevatron energies (1.96 TeV). The following relation was used to retrieve the BEC function $C_2(Q)$ from simulated data:
\begin{equation}
C_2 (Q) = N^{ \pm  \pm } (Q)/N^{ref} (Q),\label{Konstrukcia C2}
\end{equation}

\noindent where $N^{ \pm  \pm } (Q)$ is a number of like-sign pairs with four momentum difference $Q = \sqrt {(p_1-p_2)_{\mu}^2 }$, $p_{\mu_{i}}(i=1,2)$ are four-momenta of produced particles. $N^{ref} $ is the number of pairs without BEC. The reference sample can be created in two different ways. One can use the unlike-sign pairs from the same event or take the like-sign pairs from different events. Both approaches were employed in the present work.\\
The correlation procedure in PYTHIA is based on the following assumptions. First of all, the $Q_{ij}=\sqrt {(p_i  + p_j )^2  - 4m^2 }$ value is evaluated for each pairs of produced particles with the mass $m$. Then the shifted $Q'_{ij}$ is found as a solution of the following equation\\

\begin{equation}
\mathtt{\int\limits_0^{Q_{ij} } {\frac{{Q^2 dQ}}{{\sqrt {Q^2  + 4m^2 } }}}  = \int\limits_0^{Q'_{ij} } {C_2 (Q)\frac{{Q^2 dQ}}{{\sqrt {Q^2  + 4m^2 } }}}},\label{Pythia a BEC}
\end{equation}
using the chosen parameterization of $C_2(Q)$ function.
\section{Experiment ALEPH vs. $QFT_{\beta}$ model of BEC.}
\indent In general, the shape of BEC $C_2(Q)$ function is model dependent. The most simple form of Goldhaber-like parametrization for $C_2(Q)$ [8]
\begin{equation}
C_2 (Q) = \xi (N)(1 + \lambda e^{ - Q^2 R^2 } )(1 + \varepsilon Q),\label{c2 aleph}
\end{equation}
has been used for data fitting in LEP and HERA experiments.
In formula (3), $\xi(N)$ is the normalization factor, depending on multiplicity $N$, $\lambda$ is so-called the coherence strength factor, meaning $\lambda = 1$ for fully incoherent and $\lambda = 0$ for fully coherent sources; the parameter $R$ is interpreted as a radius of the particle source, assumed to be spherical in this parameterisation $[8,9]$. The linear term in (3) accounts for long-range correlations outside the region of BEC. We use the $C_2(Q)$ function within $QFT_\beta$ approach [1,2] in the form:
\begin{equation}
C_2(Q)=\xi (N)\left[ 1+\frac{{2\alpha }}{{\left( {1 + \alpha } \right)^2 }}\sqrt {\tilde \Omega \left( q \right)}+\frac{1}{{\left( {1 + \alpha } \right)^2 }}\tilde \Omega \left( q \right)\right],\label{c2 Kozlov}
\end{equation}

\noindent where the multiplicity $N$ depending factor $\xi(N)$ is equal to: 
$\xi(N) = {{\left\langle {N\left( {N - 1} \right)} \right\rangle } \mathord{\left/
 {\vphantom {{\left\langle {N\left( {N - 1} \right)} \right\rangle } {\left\langle N \right\rangle ^2 }}} \right.
 \kern-\nulldelimiterspace} {\left\langle N \right\rangle ^2 }}$.

\begin{figure}[h]
      \centering
      \epsfig{file=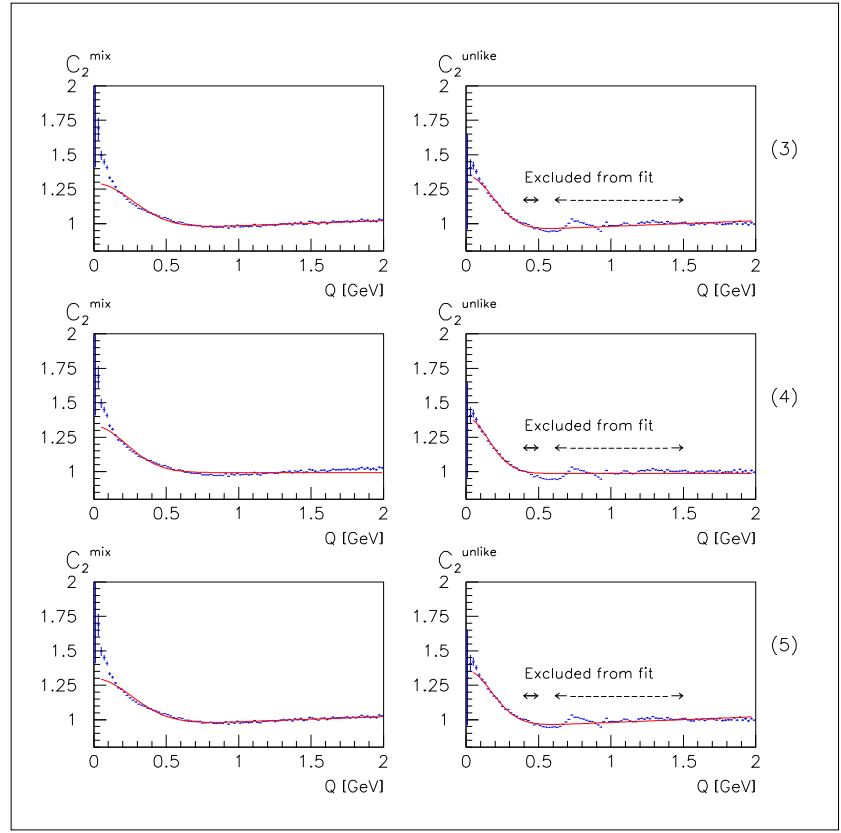,height=11cm,width=11cm,clip=}
      \caption{The one-dimensional correlation function constructed using the ALEPH data with the mixed reference sample (left), $C_2^{mix}(Q)$, and with the unlike-sign reference sample (right), $C_2^{unlike}(Q)$. The data are fitted with Eqs. (3), (4) and (5).}
\end{figure}

\begin{table}
\begin{center}
\begin{tabular}{cccc}
 &  \footnotesize $C_2^{mix}(Q)$ &\footnotesize $C_2^{+-}(Q)$ \\
\hline
\footnotesize $\xi(N)$ &\footnotesize 0.948 $\pm$ 0.001 &\footnotesize 0.936 $\pm$ 0.001 &\\
\footnotesize $\lambda$ &\footnotesize 0.362 $\pm$ 0.006 &\footnotesize 0.438 $\pm$ 0.006 &\\
\footnotesize $R(fm)$ &\footnotesize 0.528 $\pm$ 0.005 &\footnotesize 0.777 $\pm$ 0.007 & (3)\\
\footnotesize $\epsilon $ &\footnotesize (0.768 $\pm$ 0.01)10$^{-2}$ &\footnotesize (0.905 $\pm$ 0.002)10$^{-2}$ &\\
\footnotesize $\chi ^2/ndf$ &\footnotesize 513 / 94 &\footnotesize 432 / 72 &\\
\hline
\footnotesize $\xi (N)$ &\footnotesize 0.993 $\pm$  0.001 &\footnotesize 0.9871 $\pm$  0.0007 &\\
\footnotesize $\alpha $ &\footnotesize 4.32 $\pm$  0.08 &\footnotesize 3.28 $\pm$ 0.08 &\\
\footnotesize $R(fm)$ &\footnotesize 0.827 $\pm$  0.008 &\footnotesize 1.21 $\pm$  0.01 & (4)\\
\footnotesize $\chi ^2/ndf$ &\footnotesize 834.5 / 95 &\footnotesize 1108 / 75 &\\
\hline
\footnotesize $\xi (N)$ &\footnotesize 0.946 $\pm$ 0.003 &\footnotesize 0.938 $\pm$  0.002 &\\
\footnotesize $\alpha $ &\footnotesize 3.85 $\pm$ 0.07 &\footnotesize 2.92 $\pm$ 0.06 &\\
\footnotesize $R(fm)$ &\footnotesize 0.713 $\pm$ 0.009 &\footnotesize 1.04 $\pm$ 0.01 & (5)\\
\footnotesize $r_f (fm)$ &\footnotesize (0.79 $\pm$ 0.05)10$^{-2}$ &\footnotesize (0.87 $\pm$  0.04)10$^{-2}$ &\\
\footnotesize $\chi ^2/ndf$ &\footnotesize 483.7 / 94 &\footnotesize 465.8 / 74 &\\
\hline
\end{tabular}
\caption[smallcaption]{Comparison of the fit output parameters obtained with different $C_2(Q)$ functions. The ALEPH data were fitted with Eqs. (3), (4) and (5).}
\label{label} 
\end{center}
\end{table}
The important parameter $\alpha $ summarizes our knowledge of other than space-time characteristics of the particle emitting source. $\tilde \Omega(q)$ is a Fourier image of hadronizing source given by $\tilde \Omega (q) = e^{ - q^2}\gamma (n) = e^{ - Q^2 R^2}\gamma (n) $, where $\gamma (n) = \frac{{n^2 (\bar \omega )}}{{n(\omega )n(\omega ')}}$, $\bar \omega  = \frac{{\omega  + \omega '}}{2}$, and

\[
n(\omega ) \equiv  n(\omega ,\beta ) = \frac{1}{{e^{(\omega  - \mu )\beta} - 1 }},
\]
\noindent where $n(\omega,\beta )$ is the number of particles with the energy $\omega $ in the thermal bath with temperature $T = \frac{1}{\beta }$.

\noindent For a more complete analysis we use the modified formula for $C_2(Q)$:
\begin{equation}
C_2(Q) = \xi (N) \left[ 1 + \frac{{2\alpha }}{{( {1 + \alpha })^2 }}\sqrt {\tilde \Omega \left( q \right)}+ \frac{1}{{\left( {1 + \alpha } \right)^2 }}\tilde \Omega \left( q \right)\right](1 + r_fQ),\label{c2 Kozlov modified}
\end{equation}

\noindent where the consequence of the Bogolyubov's principle of weakening of correlations at large distances was not ignored; $r_f$ is the parameter of weakening of correlations.\\
\noindent The correlation functions, $C_2^{unlike}(Q)$ and $C_2^{mix}(Q)$, obtained using the ALEPH data, are shown in Fig. 1. In Table 1., we present the main retrieved parameters containing in formulae (3), (4), and (5) using the ALEPH data.

\section{PYTHIA Simulations}

\indent We fit the simulated data from PYTHIA with the most common $C_2(Q)$ function (3) at $\varepsilon =0$.
In the first step, we reproduced the parameters $\lambda$ and R, that had been put into PYTHIA at the beginning of simulation. We have found out there are non-vanishing correlations between $\pi^0$ mesons even if we turn off the BEC.Therefore in the next analysis we take into account only $\pi^+$ and $\pi^-$ mesons.\\
\indent Fig. 2 demonstrates how the output parameter $R$ behaves as function of input parameters. The green area around $R_{in}=1.4fm$ (right plots) is the area of the best reproduction of the parameters. It means, that the difference between input and output value of parameter $\Delta R= R_{out}-R_{in}$ in this area is approximately zero. The red area in right bottom side corresponds to $\Delta R \cong 0.2fm$ and the blue area in top left side to $\Delta R  \cong  - 0.5fm$. The output parameter $\lambda$ has no a good reproducing area. Each output value of the parameter $\lambda$ was smaller than the input one in the range $0.2$ up to $0.5$.\\

\begin{figure}[h]
   \centering
   \epsfig{file=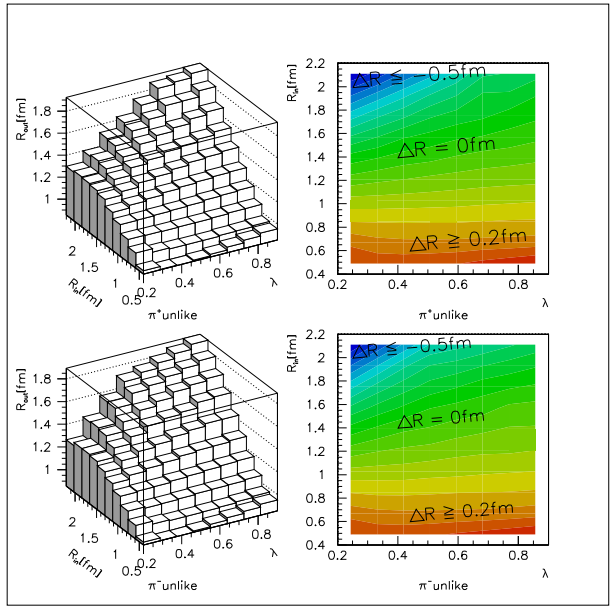,height=8cm,width=8cm,clip=}
   \caption{The dependence of the output parameter $R_{out}$ on the input parameters $R_{in}$ and $\lambda$ (left). The different colors correspond to different values of $\Delta R = R_{out}-R_{in}$ (right).}
\end{figure}

\newpage 
\section{Two-particle BEC for CDF}
\indent In the next step, we have investigated a possibility to study the two-particle correlations in the CDF experiment. We blurred all components of momenta of each particle in accord with the CDF detector resolutions [10].
The PYTHIA simulations show us, that a lot of pions have energy too small for reaching the hadron calorimeter. Therefore we get new particle energies using blurred momenta and particles masses (Fig. 3), employing, for the blurring procedure, the realistic CDF tracking system resolution.\\ 
It is evident that taking into account realistic CDF detector resolutions we are able to observe the BEC effects, although the $C_2(Q)$ function is less profound in this case.\\
In Fig 4. one can see a decrease of the output parameter $R_{out}$, taken as a function of the input parameter $R_{in}$, with increasing cut on energy. For suppression of the decrease it is therefore important to register also low energy particles.\\

\begin{figure}[h]
   \begin{minipage}[t]{.5\linewidth}
      \centering
      \epsfig{file=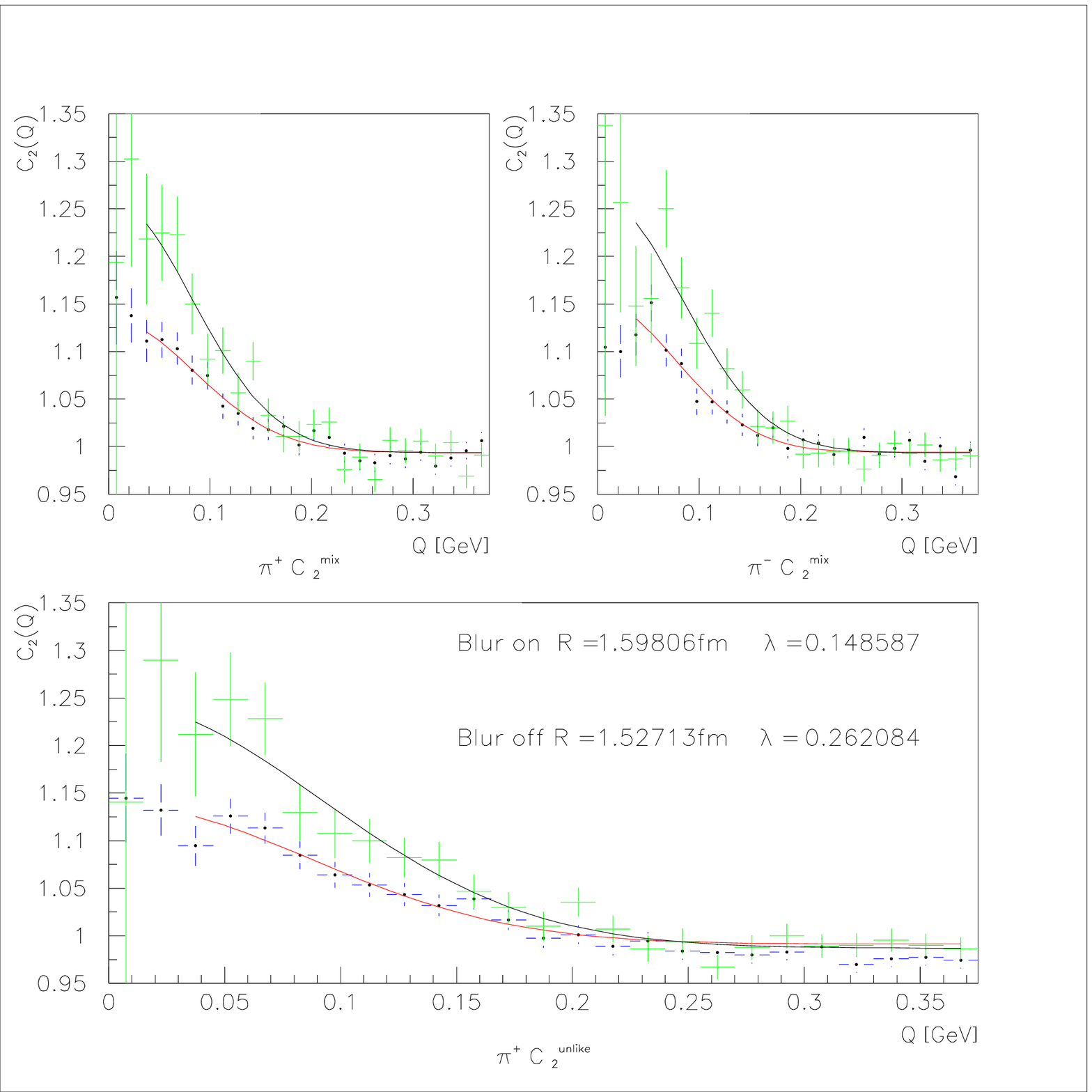,height=8cm,width=8cm,clip=}
      \caption{The dots (black curve) correspond to the $C_2(Q)$ function obtained from PYTHIA with no blurring and the squares (red curve) correspond to the $C_2(Q)$ with blurring.}
   \end{minipage}
   \hskip0.02\textwidth
   \begin{minipage}[t]{.5\linewidth}   
      \centering
      \epsfig{file=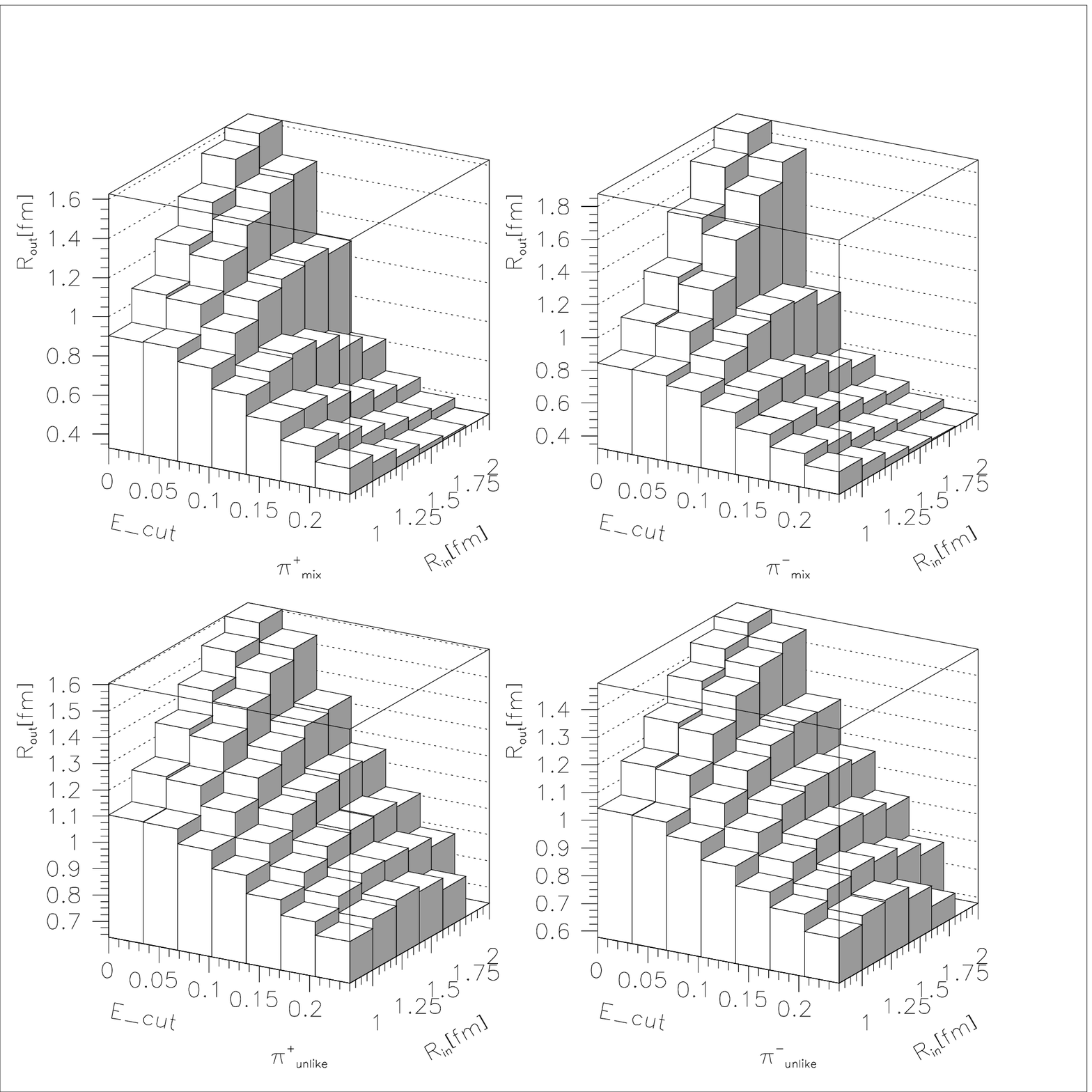,height=8cm,width=8cm,clip=}
      \caption{The dependence of output parameter $R$ on input parameter $R$ and different energy cuts; $\lambda$ is set to 0.4.}
   \end{minipage}
\end{figure}

\section{Summary}
Based on $QFT_{\beta }$ we carried out the numerical analysis of the ALEPH data applied to two-particle BEC. As a result, the important parameters of $C_2(Q)$ function $(R, \alpha,\xi(N), r_f)$ are retrieved. We investigated the possibility of the CDF detector at the Tevatron to explore the two-particle correlations in the final state. As a result, there will be possible to observe the BEC effects in case of charged pions although the $C_2(Q)$ function may be slightly suppressed. The importance for measuring of low-energy particles for BEC studies at the CDF is established.\\

\section{Acknowledgement}
\indent The present work was partly supported by EEC RTN contract HPRNT-CT-00292-2002 and by grant agency VEGA of SAS Slovakia. The authors thank to I.Aracena (University of Bern) for providing us with the ALEPH two-particle correlation data.\\

\end{document}